\documentclass[english,twocolumn]{revtex4}

\usepackage{graphicx}

\usepackage{bm}

\usepackage{cases}
\usepackage{eqnarray}

\usepackage{amssymb}

\usepackage{slashed}

\usepackage{color}

\usepackage{subfigure}
\usepackage{multirow}

\newcommand*{\affmark}[1][*]{\textsuperscript{#1}}

\begin{document}

\title{ A neoclassically optimized compact stellarator with four planar coils}

\author{Guodong Yu\affmark[1]}
\author{Zhichen Feng\affmark[1]}
\author{Peiyou Jiang\affmark[1]}
\author{Neil Pomphrey\affmark[2]}
\author{Matt Landreman\affmark[3]}




\author{GuoYong Fu\affmark[1]}\thanks{corresponding author's Email: gyfu@zju.edu.cn}

\affiliation{\affmark[1]\small{Institute for Fusion Theory and Simulation and Department of physics, Zhejiang University, Hangzhou 310027, China}}

\affiliation{\affmark[2]\small{Princeton Plasma Physics Laboratory, Princeton, New Jersey 08543, USA}}


\affiliation{\affmark[3]\small{Institute for Research in Electronics and Applied Physics, University of Maryland, College Park, Maryland, 20742, USA}}









\begin{abstract}

A neoclassically optimized compact stellarator with simple coils has been designed. The magnetic field of the new stellarator is generated by only four planar coils including two interlocking coils of elliptical shape and two circular poloidal field coils. The interlocking coil topology is the same as that of the Columbia Non-neutral Torus (CNT)\cite{Pedersen2002Confinement}.  The new configuration was obtained by minimizing the effective helical ripple \cite{Nemov1999Evaluation} directly via the shape of the two interlocking coils. The optimized compact stellarator has very low effective ripple in the plasma core implying excellent neoclassical confinement. This is confirmed by the results of the drift-kinetic code SFINCS\cite{Landreman2014Comparison}
showing that the particle diffusion coefficient of the new configuration is one order of magnitude lower than CNT's.

\end{abstract}

 \maketitle

\section{introduction}

Stellarators have enjoyed a renaissance recently as the new advanced stellarator W7-X\cite{Wolf2017Major} began to operate in 2015 and the optimized neoclassical confinement is demonstrated\cite{Dinklage2018Magnetic}. The success of W7-X has demonstrated that optimized stellarators of complicated 3D coils and considerable size can be constructed to the precision required to produce good flux surfaces, excellent neoclassical confinement and other desirable properties as designed. It can be argued that advanced stellarators now form a main line approach to magnetic fusion energy. They have
advantages over tokamaks of natural steady state operation without disruptions. The magnetic field of stellarators is mainly provided by external coils. Therefore the physics properties of stellarators can be largely controlled by external coils and can thus be optimized by varying coil geometry. However 3D stellarator coils usually have complex 3D geometry and they are difficult and costly to build. It is thus important to explore the possibilities of optimized stellarators with simple coils.

In this work a neoclassically optimised period $2$ compact stellarator with only four simple coils has been designed as a candidate for a low-cost toroidal
magnetic confinement device at Zhejang University. The new stellarator is similar to the Columbia Non-neutral Torus (CNT)\cite{Pedersen2002Confinement} with two InterLocking (IL) coils and two poloidal field coils. The new configuration is obtained by direct optimization of the shape of the two interlocking coils. This direct method is different from the conventional two-stage optimization where the first stage is optimization of physics properties of stellarators from the shape of the last closed flux surface. The second stage is design of 3D coil set which is optimized in such a way that the shape of plasma boundary it generates closely matches the plasma boundary obtained in the first stage. The two-stage method usually works well and it was successful in design of advanced stellarators such as HSX\cite{Canik2007Experimental}, W7-X\cite{Dinklage2018Magnetic}, NCSX\cite{NELSON2003Design}, ESTELL\cite{Drevlak2013ESTELL}, CFQS\cite{LIU2018Magnetic}, and a new design of quasi-axisymmetric stellarator\cite{Henneberg2019Properties}. However it suffers from the fact that the coils  found in the second stage cannot perfectly recover the optimized plasma boundary obtained in the first stage. Thus usually some iterations between the first stage and second stage are needed in order to obtain the desired physics and engineering properties. In view of this, we adopted
the direct optimization method from coils. Specifically we carry out optimization by varying the shape of stellarator coils to directly control the physics
properties of vacuum magnetic field of stellarators. Our primary optimization target is the so called $1/\nu$ neoclassical transport\cite{Nemov1999Evaluation} due to helical ripple. This transport is due to finite magnetic drift of trapped particles in helical wells. The neoclassical  transport is a serious issue for stellarators. The $1/\nu$ scaling is very unfavorable for fusion reactors where the plasma temperature is necessarily high and the collision frequency is very low.

We chose compact stellarator topology of CNT type in our design for two reasons. First, CNT is arguably the world's simplest stellarator with only four circular coils including two interlocking coils and two poloidal field coils. Therefore it is relatively easy to build. Second, our direct optimization method is suitable for stellarator design of CNT type because the shape of only one coil needs to be considered in the optimization and thus the number of degrees of freedom is modest. It should be pointed out that the original goal of CNT was non-neutral plasma experiment and thus the neoclassical transport due to helical ripple was not emphasized. In contrast, our design goal is experimental study of fully ionized neutral plasmas. Therefore the neoclassical confinement is our primary focus in the configuration optimization. We will show that excellent neoclassical confinement in the plasma core can be achieved with two interlocking coils of elliptical planar shape. The optimized configuration is to be called Zhejiang university Compact Stellarator (ZCS).

The paper is organized as following. Section \uppercase\expandafter{\romannumeral2} describes the detailed optimization process.  An optimized configuration with good neoclassical confinement is described including coil geometry and magnetic flux surfaces as well as rotational transform profile. Section \uppercase\expandafter{\romannumeral3} shows the calculated $1/\nu$ neoclassical transport coefficient $\epsilon_{eff}^{3/2}$ of ZCS. In section \uppercase\expandafter{\romannumeral4}, the results of neoclassical transport obtained with the drift-kinetic code SFINCS\cite{Landreman2014Comparison} are presented and discussed. In section \uppercase\expandafter{\romannumeral5}, the effects of finite plasma beta on equilibrium and the effective ripple are studied. In section \uppercase\expandafter{\romannumeral6}, conclusions of this work are given.

\section{Optimization methods}

For the purpose of carrying out stellarator optimization directly from coils, a code suite has been developed for calculating vacuum magnetic field from coils, magnetic flux surfaces as well as particle motions in the magnetic field. The magnetic field is calculated from current-carrying coils straight forwardly using the Biot-Savart law. A line current is assumed for simplicity. The vacuum magnetic flux surfaces and corresponding rotational transform profile are calculated by following the magnetic field lines. The $1/\nu$ neoclassical coefficient is calculated by integrating along magnetic field lines\cite{Nemov1999Evaluation}. The neoclassical transport is also evaluated by the drift kinetic code SFINCS\cite{Landreman2014Comparison}.

Here we describe the method used for optimizing stellarators directly from coils. In this work our main goal is optimization of neoclassical confinement in the $1/\nu$ regime by minimizing the effective ripple coefficient $\epsilon_{eff}^{3/2}$.

Assuming every coil is a closed smooth curve, we use Fourier series\cite{Zhu2017New} to define the shape of each coil in Cartesian coordinate as

\begin{numcases}{}
x=x_{c,0}+\sum_{n=1}^{n_f}[x_{c,n}\cos(nt)+x_{s,n}\sin(nt)],\\
y=y_{c,0}+\sum_{n=1}^{n_f}[y_{c,n}\cos(nt)+y_{s,n}\sin(nt)],\\
z=z_{c,0}+\sum_{n=1}^{n_f}[z_{c,n}\cos(nt)+z_{s,n}\sin(nt)],
\end{numcases}

where the angle parameter $t$ ranges $[0,2\pi]$ so that the coil curve is closed. From above formula we see that the shape of each coil is determined by $3\times(2n_f+1)$ Fourier harmonics, with $n_f$ being the cutoff harmonic number. It should be noted that, almost any closed smooth curve without straight sections or sharp corners can be represented well by Fourier harmonics.

As mentioned above, we choose CNT as starting point of our optimization. Our approach is optimizing neoclassical confinement by varying the shapes of the two interlocking coils. Furthermore, because it is a two period stellarator, the shapes of the two interlocking coils are necessarily the same. Therefore the degree of freedom is minimized and is much smaller than that of conventional stellarators. As in CNT, we
incorporate a pair of up-down symmetric circular poloidal field (PF) coils whose main
role is to provide a vertical field. The distance between these coils is fixed, as is their
current. Only the ratio of their current $I_{PF}$ to the that of the interlocking coils is
important.

\begin{figure}

\includegraphics[scale=0.4]{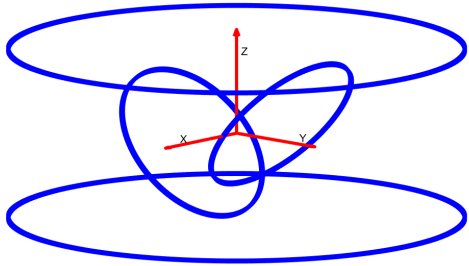}
\caption{CNT coils}\label{fig1}

\end{figure}

FIG.\ref{fig1} plots the coil
configuration of CNT with the machine centerline chosen to lie along the axis. Since
the PF coils are axisymmetric, we are free to fix the orientation of IL coils so that its semi-major
axes points along the $x$ axis and its semi-minor axes lies in the $y-z$ plane.
The two IL coils can therefore be represented by only four non-zero Fourier
harmonics including $x_{c,0}=\pm0.313, x_{c,1}=\pm0.405, y_{s,1}=\mp0.255, z_{s,1}=0.315$, in which $x_{c,0}$  and $|x_{c,1}|=\sqrt{y_{s,1}^2+z_{s,1}^2}$ represent the half distance between the two coil centers and the radius of the coils, respectively. The shapes of coils are elliptic when $|x_{c,1}|\neq \sqrt{y_{s,1}^2+z_{s,1}^2}$ . The angle between the two coil planes is controlled by $\theta=2*\arctan{(z_{s,1}/y_{s,1})}$. If other higher order Fourier harmonics $(n>1)$ are kept, the shape of interlocking coils changes from planar coil to three dimensional coil. For poloidal field coils, Fourier harmonics $y_{s,1}=x_{c,1}=1.08$ and $z_{c,0}=\pm0.405$ represent the coil radius and the half distance between the center of two coils, respectively.

A combination of global optimization and Levenberg-Marquardt\cite{Marquardt1963An} algorithm is adopted in our optimization process. For the global optimization method, we chose an appropriate parameter range and associated $n_i$ discrete grid points for each Fourier coefficient. Thus, the total mesh points in the multi-dimensional parameter space of  all Fourier coefficients is $n_i^{N_F}$ with $N_F=3\times(2n_f+1)+2$ being the total degree of freedom. Here $N_F$ includes the current of the two IL coils and the radius of the two poloidal field coils.
Imposing the stellarator symmetry, the degree of freedom is reduced to $N_F=3\times(n_f+1)$. Each mesh point represents one unique stellarator configuration and the corresponding combined target function is evaluated. In this way, a global minimum can be found as long as the total number of mesh points are limited and the required computational resource and time is reasonable. For the case of only n=0 and n=1 harmonics are included, the total degree of freedom is only $N_F =6$ and the total number of mesh points is $10^6$ for $n_i=10$. Once a global minimum is found, we can then do a refined local search near the neighborhood of this global minimum using the Levenberg-Marquardt algorithm. In this work we focus on the globally optimized configuration with Fourier harmonics up to $n_f=1$ and the specific parameter ranges are shown in Table \ref{table1}. In this case the shape of IL coils is simply planar. It turns out that inclusion of $n \textgreater 1$ harmonics only leads to a slight improvement in the target function. Thus higher harmonics of $n \textgreater 1$ are not considered in this work.

The process of optimization is quite simple. Each mesh point in the 6 dimensional parameter space corresponds to a unique configuration. For each configuration, the position of magnetic axis is determined first from field line tracing. Then the last closed magnetic surface is located and the corresponding effective ripple coefficient
$\epsilon_{eff}^{3/2}$ is calculated.  In this way, a global minimum can be found straightforwardly.

\begin{table}

\begin{tabular}{|c|c|c|c|lll}
\cline{1-4}
                                                                                & Parameter         & Value & Range        &  &  &  \\ \cline{1-4}
\multirow{4}{*}{\begin{tabular}[c]{@{}c@{}}Interlocking \\ coil\end{tabular}}   & $x_{c,0}$         & 0.313 & 0.3$\sim$0.5 &  &  &  \\ \cline{2-4}
                                                                                & $x_{c,1}$         & 0.405 & 0.2$\sim$0.6 &  &  &  \\ \cline{2-4}
                                                                                & $y_{s,1}$         & 0.255 & 0.2$\sim$0.4 &  &  &  \\ \cline{2-4}
                                                                                & $z_{s,1}$         & 0.315 & 0.2$\sim$0.6 &  &  &  \\ \cline{1-4}
\multirow{2}{*}{\begin{tabular}[c]{@{}c@{}}Poloidal \\ field coil\end{tabular}} & $y_{s,1}=x_{c,1}$ & 1.08  & 0.5$\sim$1.2 &  &  &  \\ \cline{2-4}
                                                                                & $z_{c,0}$         & 0.405 & 0.405        &  &  &  \\ \cline{1-4}
Current ratio                                                                   & $I_{IL}/I_{PF}$    & 2.25  & 1$\sim$5     &  &  &  \\ \cline{1-4}
\end{tabular}
\caption{The specific parameters variation in optimization}
\label{table1}
\end{table}

\section{Basic parameters of new configuration ZCS}

\begin{figure}[h]

\subfigure[3D schematic]{

\label{fig2.a}

\includegraphics[scale=0.2]{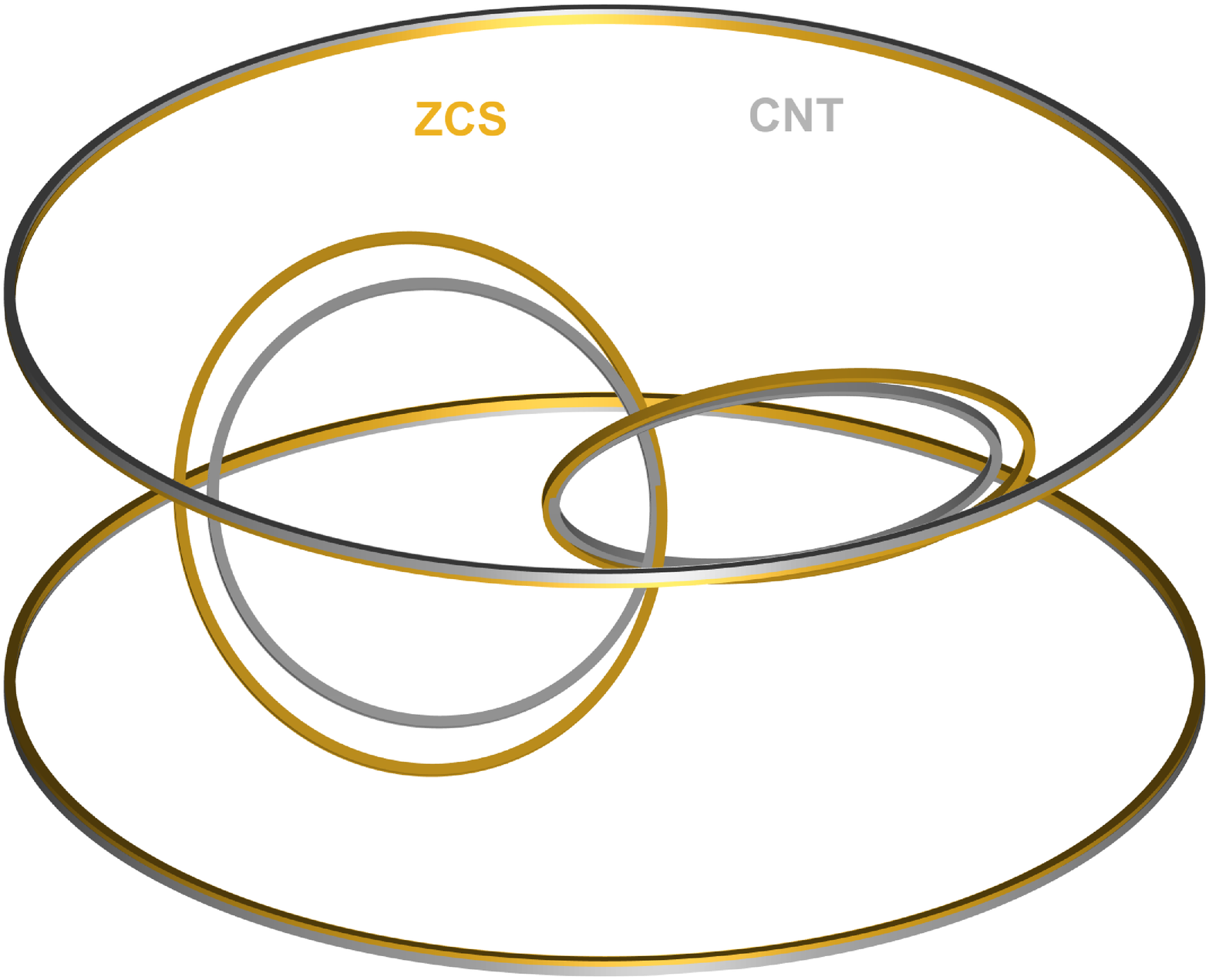}}

\subfigure[Top view ]{

\label{fig2.b}

\includegraphics[scale=0.25]{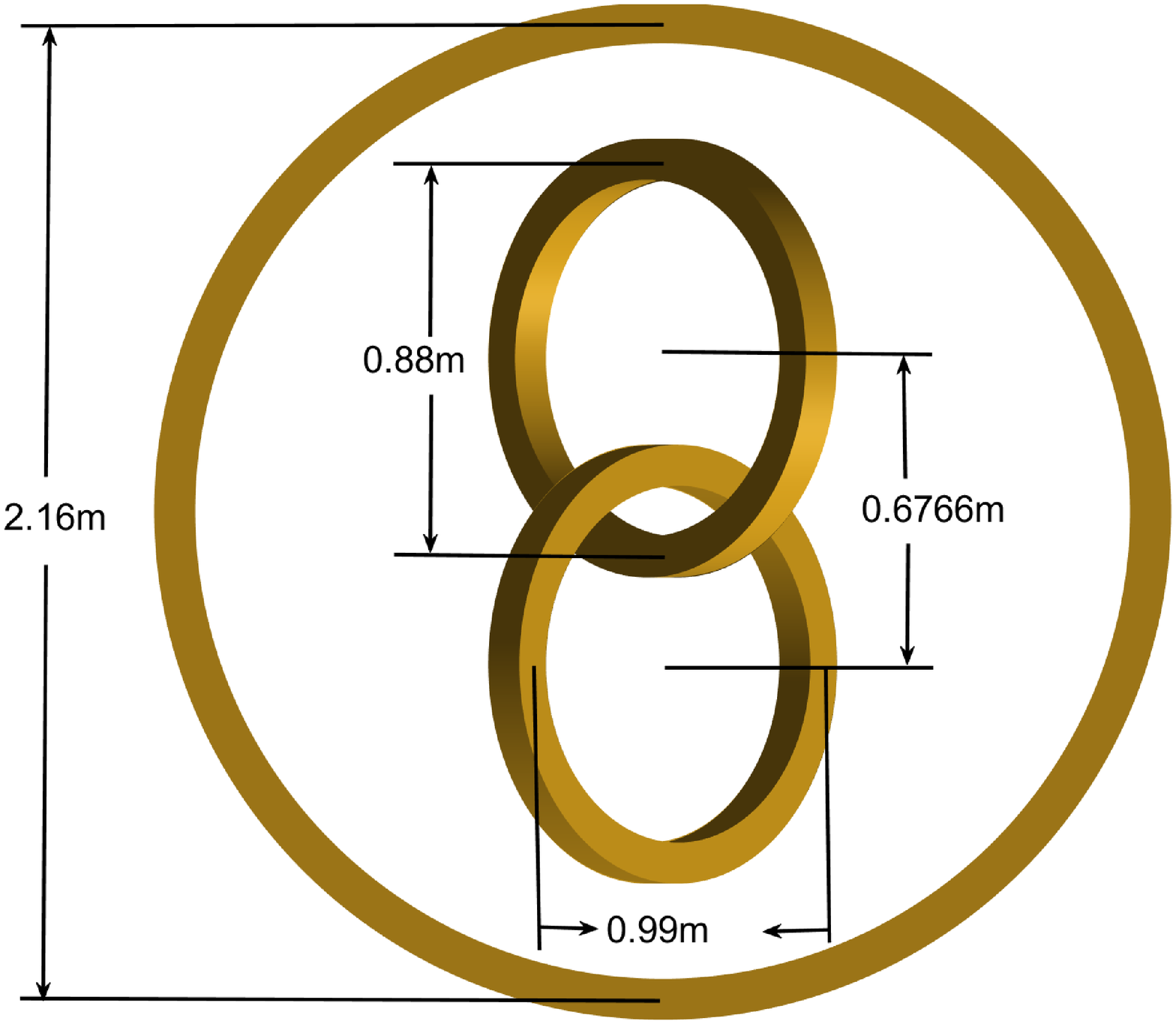}}
\subfigure[Side view ]{

\label{fig2.c}

\includegraphics[scale=0.1]{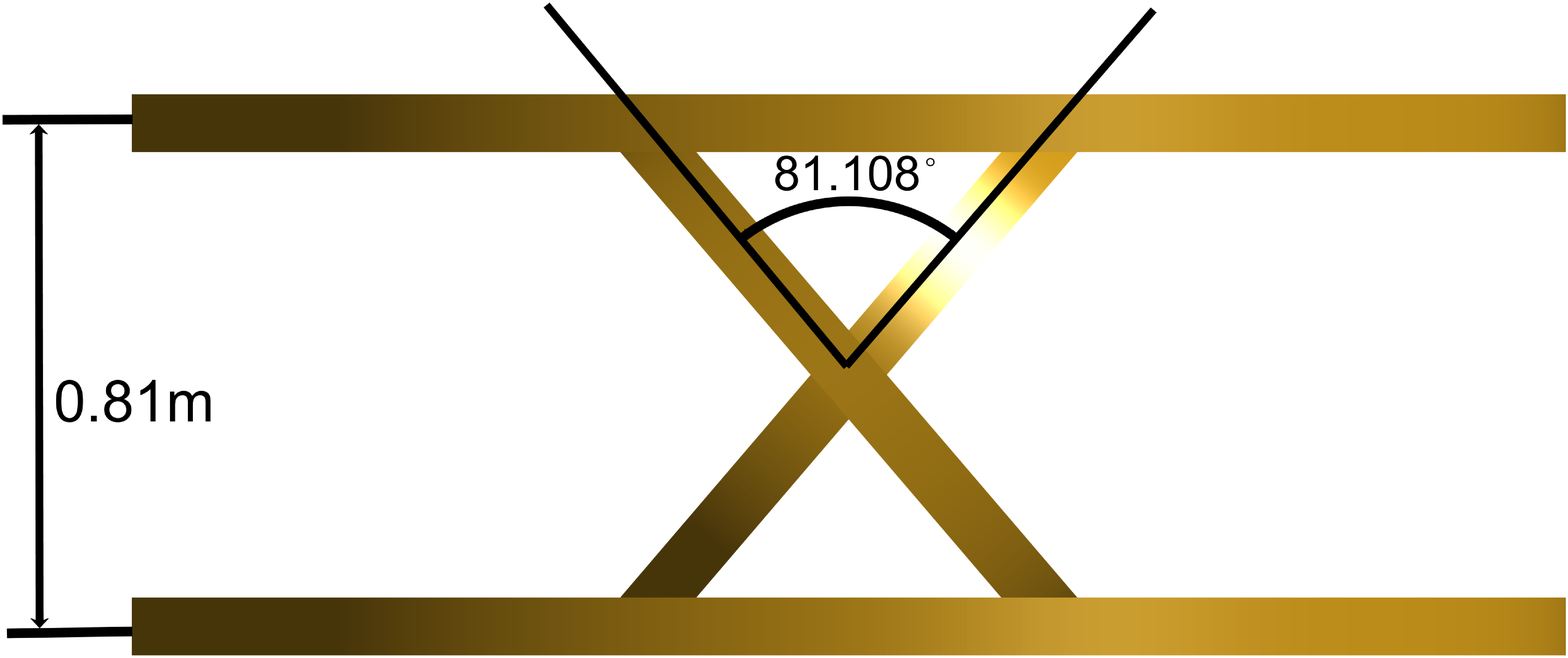}}

\caption{View of the new configuration}\label{fig2}

\end{figure}

FIG.\ref{fig2} shows the coil system of the new configuration ZCS (orange color) obtained using global optimization with only $n=0$ and $n=1$ Fourier harmonics. The IL coils of CNT (grey color) are also shown in FIG.\ref{fig2.a} for comparison. Table \ref{table2} lists coil parameters of
ZCS and CNT including Fourier coefficients of the interlocking coils and coil current ratio between IL coils and vertical field coils. The shape of IL
coils of the new configuration is now elliptical instead of circular shape of CNT's. The long and short diameter is $0.99m$ and $0.88m$ respectively (FIG.\ref{fig2.b}).
The angle and center distance between the two IL coils are $81.108^\circ$ and $0.6766m$ (FIG.\ref{fig2.c}). The radius of the circular poloidal field (PF)
coils and the center distance is $1.08m$ and $0.81m$ respectively. These two parameters are the same as CNT's. The current ratio between IL coils and PF coils is $1.6:1.0$. The main
difference between ZCS's coils and CNT's is the shape of the interlocking coils. As a result, the neoclassical confinement of the new configuration is much improved over CNT's.

\begin{table}[]
\begin{tabular}{|c|c|c|c|lll}
\cline{1-4}
                                                                                & Parameter         & CNT   & ZCS    &  &  &  \\ \cline{1-4}
\multirow{4}{*}{\begin{tabular}[c]{@{}c@{}}Interlocking \\ coil\end{tabular}}   & $x_{c,0}$         & 0.313 & 0.3383 &  &  &  \\ \cline{2-4}
                                                                                & $x_{c,1}$         & 0.405 & 0.44   &  &  &  \\ \cline{2-4}
                                                                                & $y_{s,1}$         & 0.255 & 0.322  &  &  &  \\ \cline{2-4}
                                                                                & $z_{s,1}$         & 0.315 & 0.376  &  &  &  \\ \cline{1-4}
\multirow{2}{*}{\begin{tabular}[c]{@{}c@{}}Poloidal \\ field coil\end{tabular}} & $y_{s,1}=x_{c,1}$ & 1.08  & 1.08   &  &  &  \\ \cline{2-4}
                                                                                & $z_{c,0}$         & 0.405 & 0.405  &  &  &  \\ \cline{1-4}
Current ratio                                                                   & $I_{IL}/I_{PF}$    & 2.25  & 1.6    &  &  &  \\ \cline{1-4}
Major radius                                                                    & $R(m)            $  & 2.4  & 2.0    &  &  &  \\ \cline{1-4}
Minor radius                                                                    & $a(m)            $  & 0.10  & 0.10    &  &  &  \\ \cline{1-4}
Aspect  ratio                                                                   & $R/a       $       & 2.4  &2.0     &  &  &  \\ \cline{1-4}
Volume                                                                          & $V(m^{3})     $      & 0.049  &0.049     &  &  &  \\ \cline{1-4}
\end{tabular}
\caption{The Fourier Harmonics of ZCS in comparison with CNT}
\label{table2}
\end{table}

\begin{figure}
\includegraphics[scale=0.2]{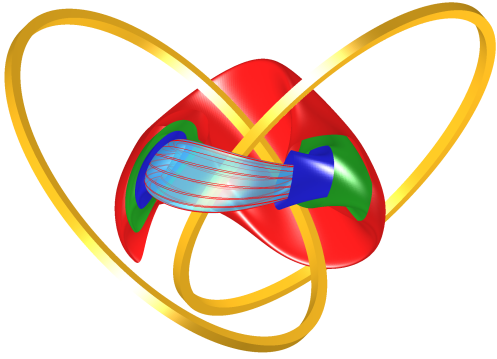}
\caption{The 3-D magnetic surface construction}\label{fig3}
\end{figure}

\begin{figure}
\includegraphics[scale=0.25]{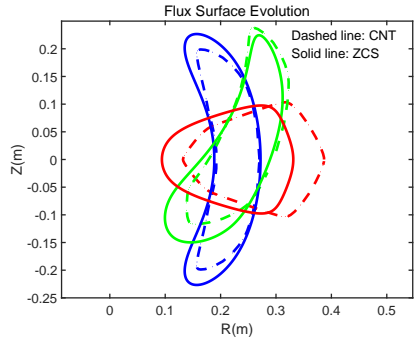}
\caption{Cross-sections of the boundary magnetic surface}\label{fig4}
\end{figure}

\begin{figure}[h]
\centering

\subfigure[ZCS-$\phi=0^\circ$]{
\begin{minipage}[t]{0.25\linewidth}
\centering
\includegraphics[scale=0.25]{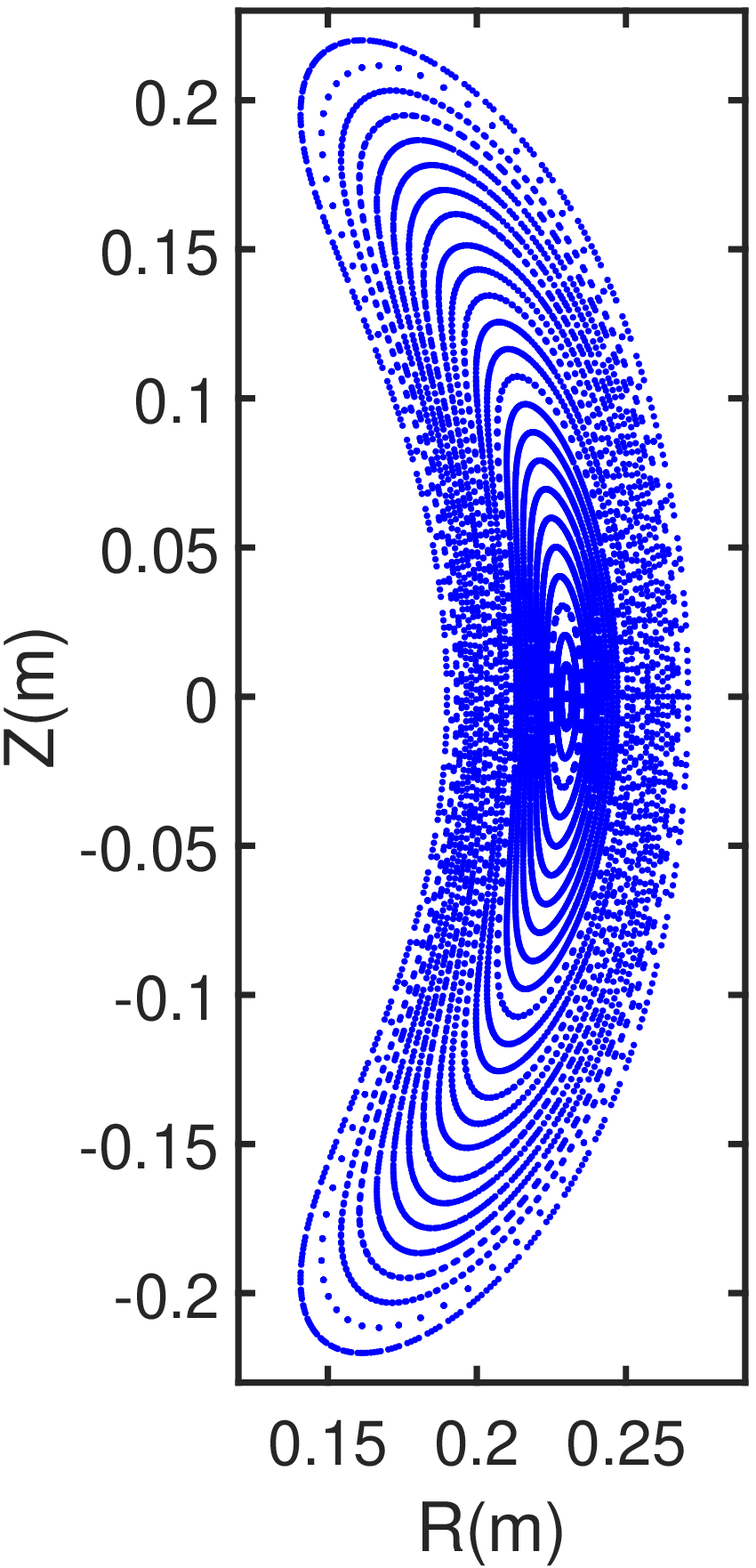}
\end{minipage}
}
\subfigure[ ZCS-$\phi=90^\circ$]{
\begin{minipage}[t]{0.25\linewidth}
\centering
\includegraphics[scale=0.25]{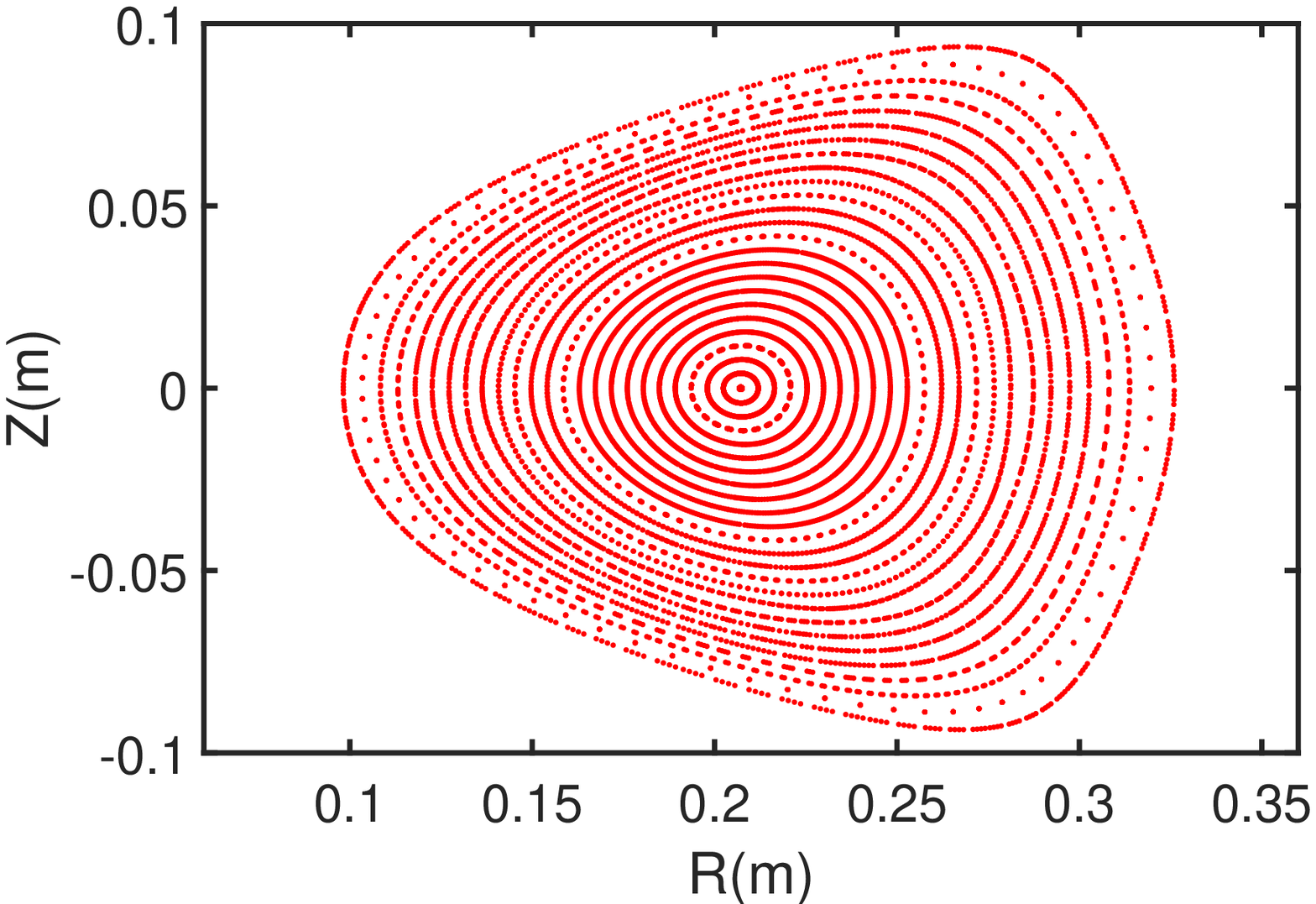}
\end{minipage}
}

\subfigure[CNT-$\phi=0^\circ$]{
\begin{minipage}[t]{0.25\linewidth}
\centering
\includegraphics[scale=0.25]{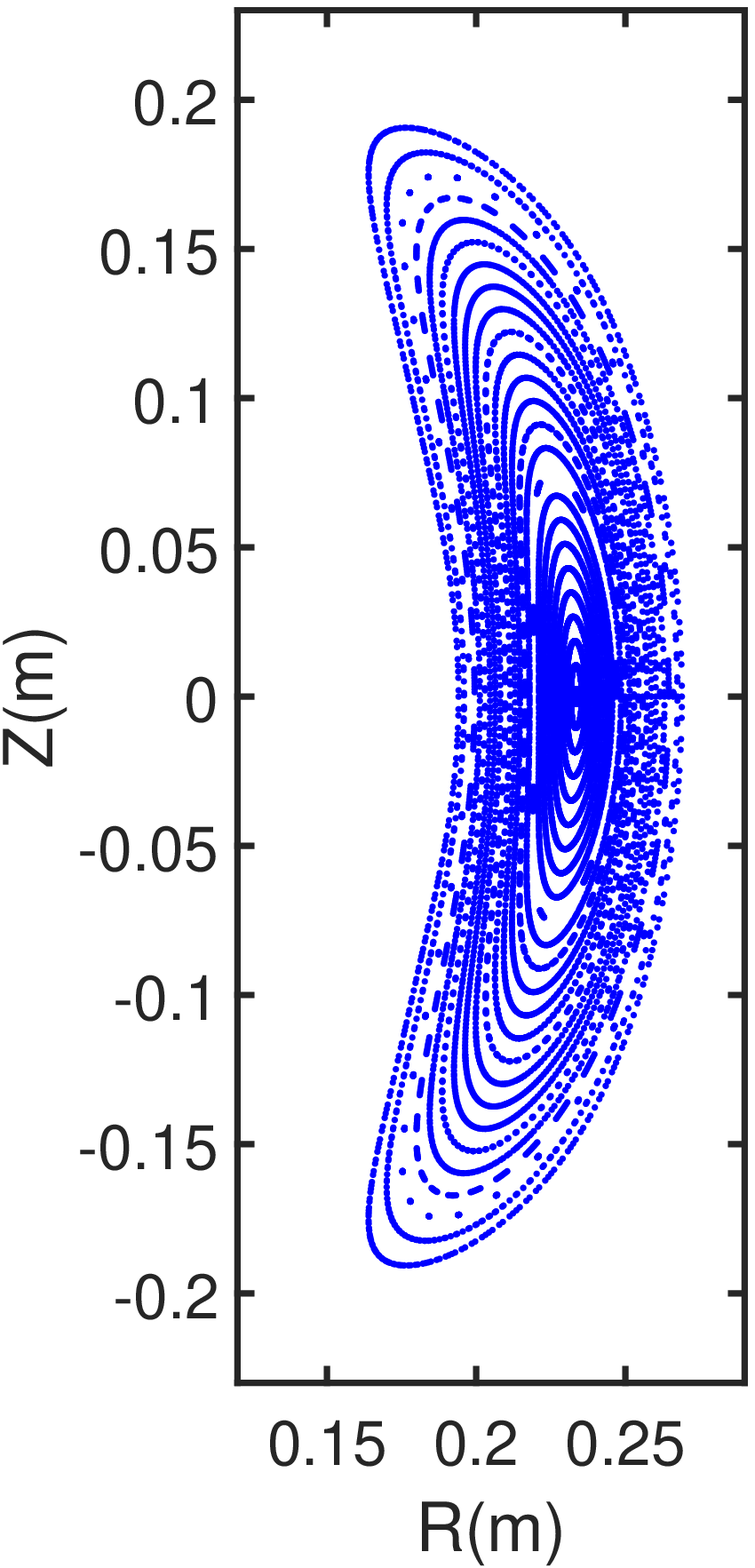}
\end{minipage}
}
\subfigure[CNT-$\phi=90^\circ$]{
\begin{minipage}[t]{0.25\linewidth}
\centering
\includegraphics[scale=0.25]{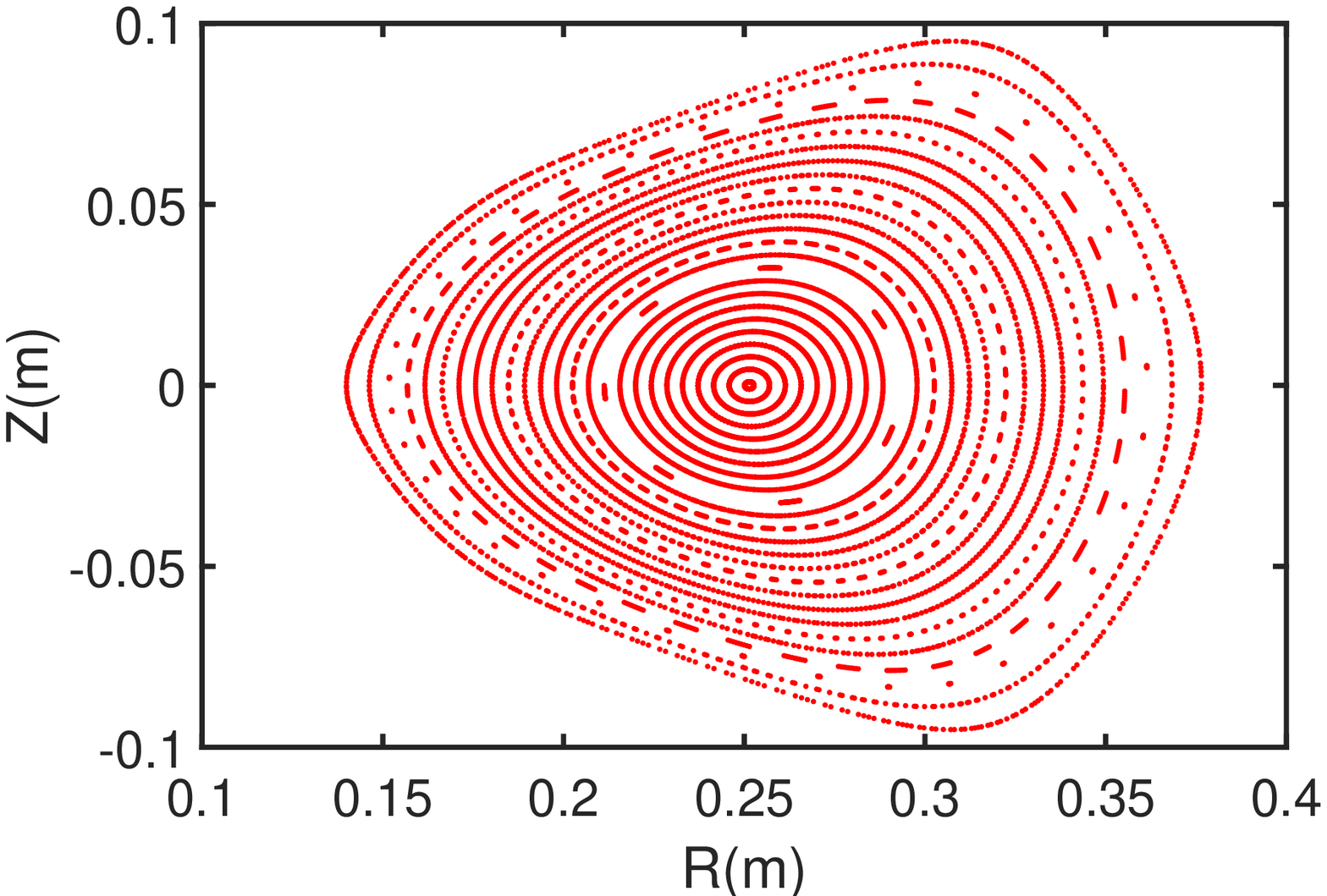}
\end{minipage}
}

\caption{Poincare plots of magnetic surfaces for vacuum }\label{fig5}

\end{figure}

FIG.\ref{fig3} plots the 3D magnetic flux surfaces relative to the two IL coils of ZCS. FIG.\ref{fig4} plots the cross sections of last closed flux surfaces of ZCS (solid lines) and CNT (dashed lines) respectively for three toroidal angles, $\phi=0^\circ, 45^\circ$ and $90^\circ$. We observe that the shapes of flux surfaces of the new configuration are similar to those of CNT's. A notable difference is that the last closed surface at $\phi=\pi/2$ shifts inward considerably as compared to that of CNT. FIG.\ref{fig5} shows the Poincare plots of magnetic surfaces of ZCS and CNT. We see that both ZCS and CNT have nice flux surfaces throughout the whole volume within the last closed surface.

\begin{figure}
\includegraphics[scale=0.25]{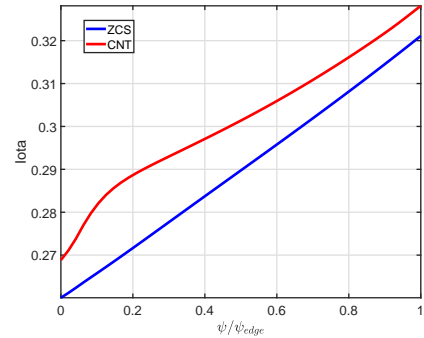}
\caption{Rotation transform }\label{fig6}
\end{figure}

The rotational transform profile of the new configuration is plotted in FIG.\ref{fig6}. We observe that the new profile is close to that of CNT.

\section{The neoclassical confinement of ZCS}

Here we show that the neoclassical confinement of the optimized compact stellarator is much better than that of CNT. The neoclassical transport is evaluated by calculating the effective ripple parameter and by using the drift-kinetic code SFINCS.

\subsection{The effective ripple}

\begin{figure}
\includegraphics[scale=0.25]{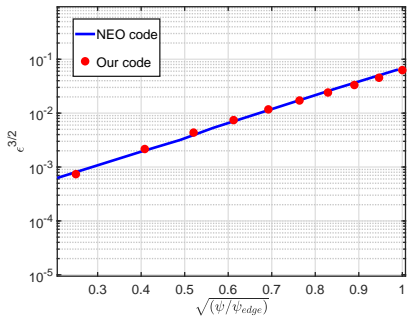}
\caption{ $\epsilon_{eff}^{3/2}$ comparison between the NEO code and our code}\label{fig7}
\end{figure}

For stellarators, the neoclassical transport due to helical ripple is a serious problem since this transport scales as $1/\nu$ for small collision frequency $\nu$. This scaling is very unfavorable for fusion reactors where plasma temperatures are necessary high and collision frequencies are small. Thus this neoclassical transport  must be minimized to achieve high plasma confinement. This $1/\nu$ transport is proportional to the effective helical ripple parameter $\epsilon_{eff}^{3/2}$. Thus this ripple parameter is chosen as the main target for our optimization. It was shown that $\epsilon_{eff}^{3/2}$ is only a function of magnetic field geometry and can be calculated straight forwardly by integrating along a magnetic field line\cite{Nemov1999Evaluation}.
A module has been developed in our code suite for calculating $\epsilon_{eff}^{3/2}$ and has been benchmarked against the NEO code\cite{STELLOPT}.
FIG.\ref{fig7} compares the calculated effective ripple of a stellarator using our module with that of the NEO code. The agreement between the results of two codes is excellent.

\begin{figure}
\includegraphics[scale=0.25]{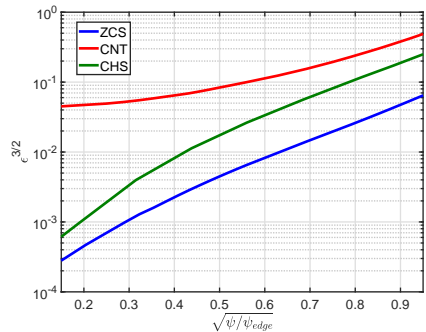}
\caption{Parameter $\epsilon_{eff}^{3/2}$ for ZCS, CNT and CHS( the data of CHS comes from \cite{Nemov2003Effective})}\label{fig8}
\end{figure}

FIG.\ref{fig8}  shows the calculated $\epsilon_{eff}^{3/2}$ of ZCS , CNT and the Compact Helical System (CHS) as a function of the normalized radial variable $\sqrt{\psi/\psi_{edge}}$ , in which $\psi_{edge}$ is the boundary poloidal flux. We observe that the effective ripple of the optimized configuration ZCS is much smaller than that of CNT especially in the core where it is two orders of magnitude smaller. Furthermore, the effective ripple of ZCS is also substantially lower than that of CHS. It is remarkable that this huge improvement in neoclassical confinement is achieved by simple planar coils of elliptical shape.

\subsection{Evaluation of neoclassical transport using SFINCS}

\begin{figure}
\subfigure[Initial density profile in SFINCS]{
\label{fig9.a}
\includegraphics[scale=0.25]{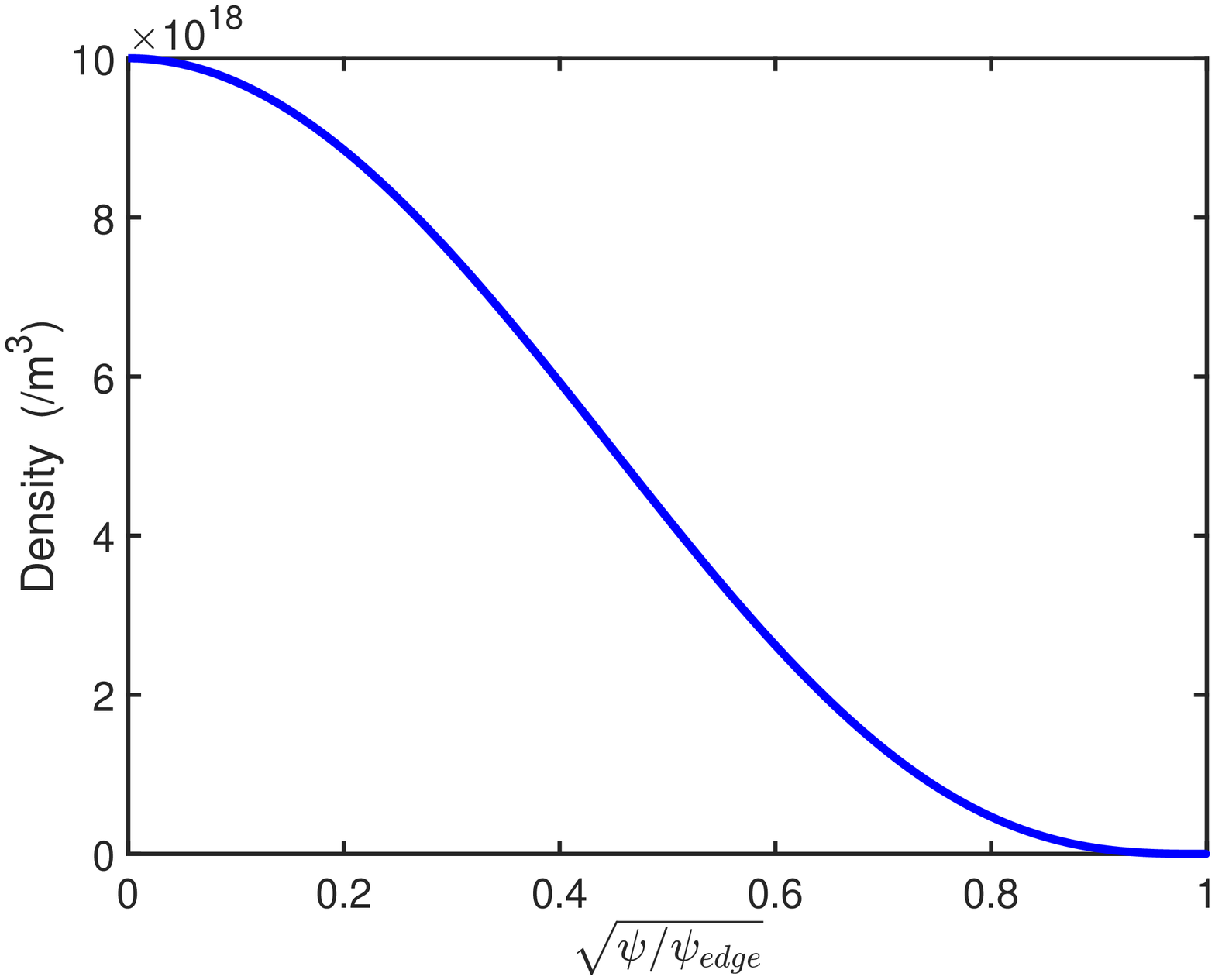}}
\subfigure[Electron transport flux density]{
\label{fig9.b}
\includegraphics[scale=0.25]{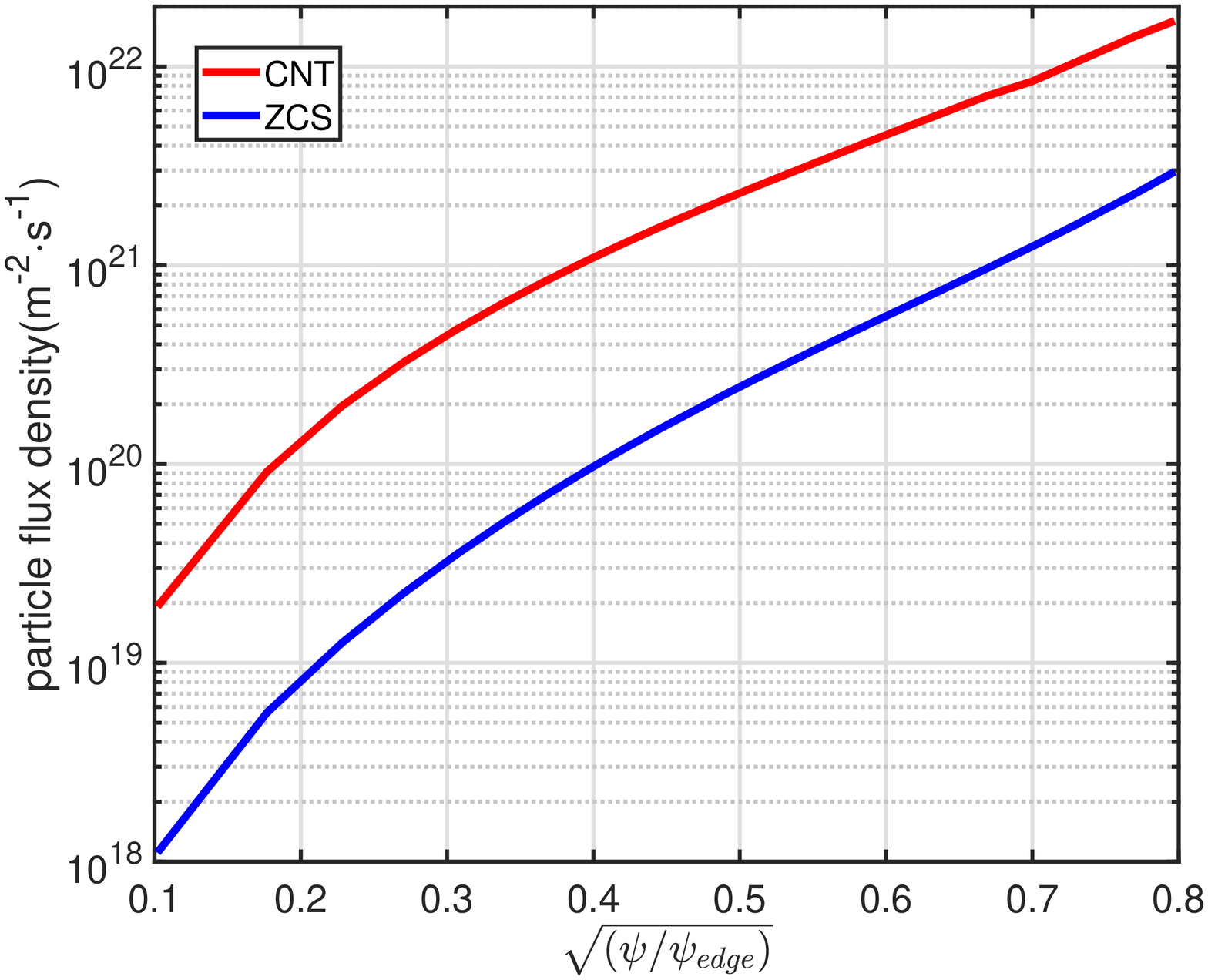}}

\caption{ SFINCS results for initial density profiles }\label{fig9}

\end{figure}

 SFINCS is a kinetic code for calculation of neoclassical transport in stellarators by solving the steady-state drift-kinetic equation for multiple species. We use SFINCS code to calculate neoclassical particle fluxes of both electrons and ions with effects of ambipolar radial electric field. The density profiles of ions and electrons considered are shown in FIG.\ref{fig9.a}. The temperature profiles are chosen to be uniform at $T_e=2T_i=200eV$ for simplicity. FIG.\ref{fig9.b} shows the electron particle fluxes of both CNT and ZCS. The results indicate that the neoclassical transport of ZCS is much lower than that of CNT especially in the plasma core where the particle flux of ZCS is one order of magnitude lower. Based on these findings, we conclude that the new optimized configuration has very good neoclassical confinement.

\subsection{Quasi-symmetry and quasi-omnigeneity}

We now consider the degree of quasi-symmetry and quasi-omnigeneity to understand the reason of good neoclassical confinement of the optimized configuration ZCS.  Quasi-symmetry is an effective concept  for improving neoclassical transport in stellarators.  Boozer showed that the particle drift orbits in stellarator are equivalent to those of axi-symmetric tokamaks if the magnetic field strength is axis-symmetric in Boozer coordinates, even though the  structure of magnetic field  is three dimensional\cite{Boozer1981Plasma}. The magnetic field distribution on a flux surface can be expressed by $B=\sum_{m,n}B_{m,n}\cos(m\theta-n\zeta)$, where $\theta$ and $\zeta$ are boozer coordinates. For quasi helical-symmetry configuration, such as Helical Symmetric Experiment(HSX)\cite{Canik2007Experimental}, the dominant Fourier components $B_{m,n}$ have a single helicity and other components are very small. For quasi axi-symmetry configurations,  the magnetic field spectrum is nearly axi-symmetric with all the  non-axi-symmetry components being very small. Quasi-axisymmetry has been used to design compact stellarators with excellent neoclassical confinement. Examples of quasi-axisymmetric stellarators include the National Compact Stellarator Experiment (NCSX)\cite{NELSON2003Design}, ESTELL\cite{Drevlak2013ESTELL} and CFQS\cite{LIU2018Magnetic}. Another approach of optimizing neoclassical confinement is quasi-omnigeneity. This approach was used to design Wendelstein 7-X (W7-X)\cite{Dinklage2018Magnetic} by minimizing the averaged particle drift.

\begin{figure}
\subfigure[CNT]{
\label{fig10.a}
\includegraphics[scale=0.25]{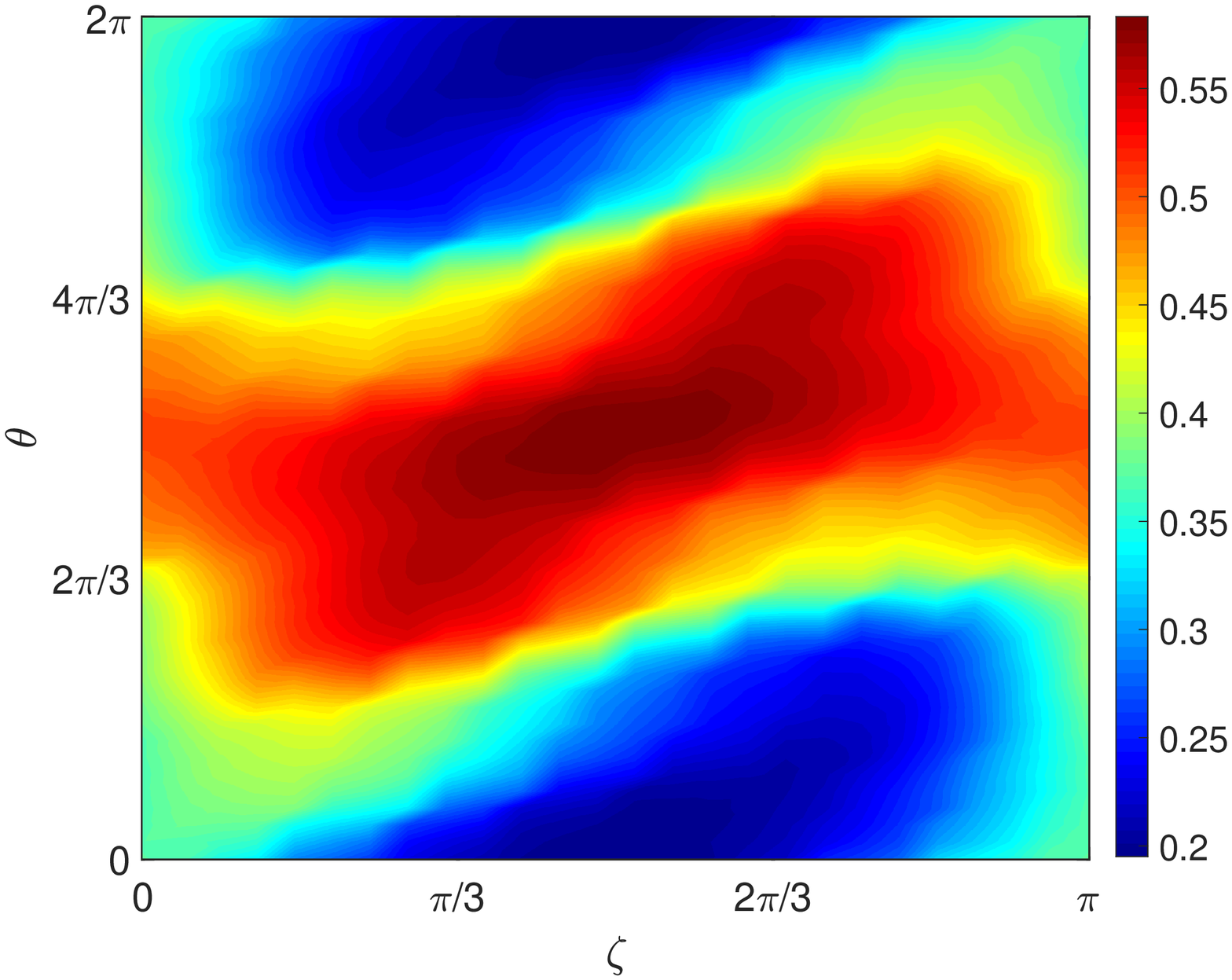}}
\subfigure[ZCS]{
\label{fig10.b}

\includegraphics[scale=0.25]{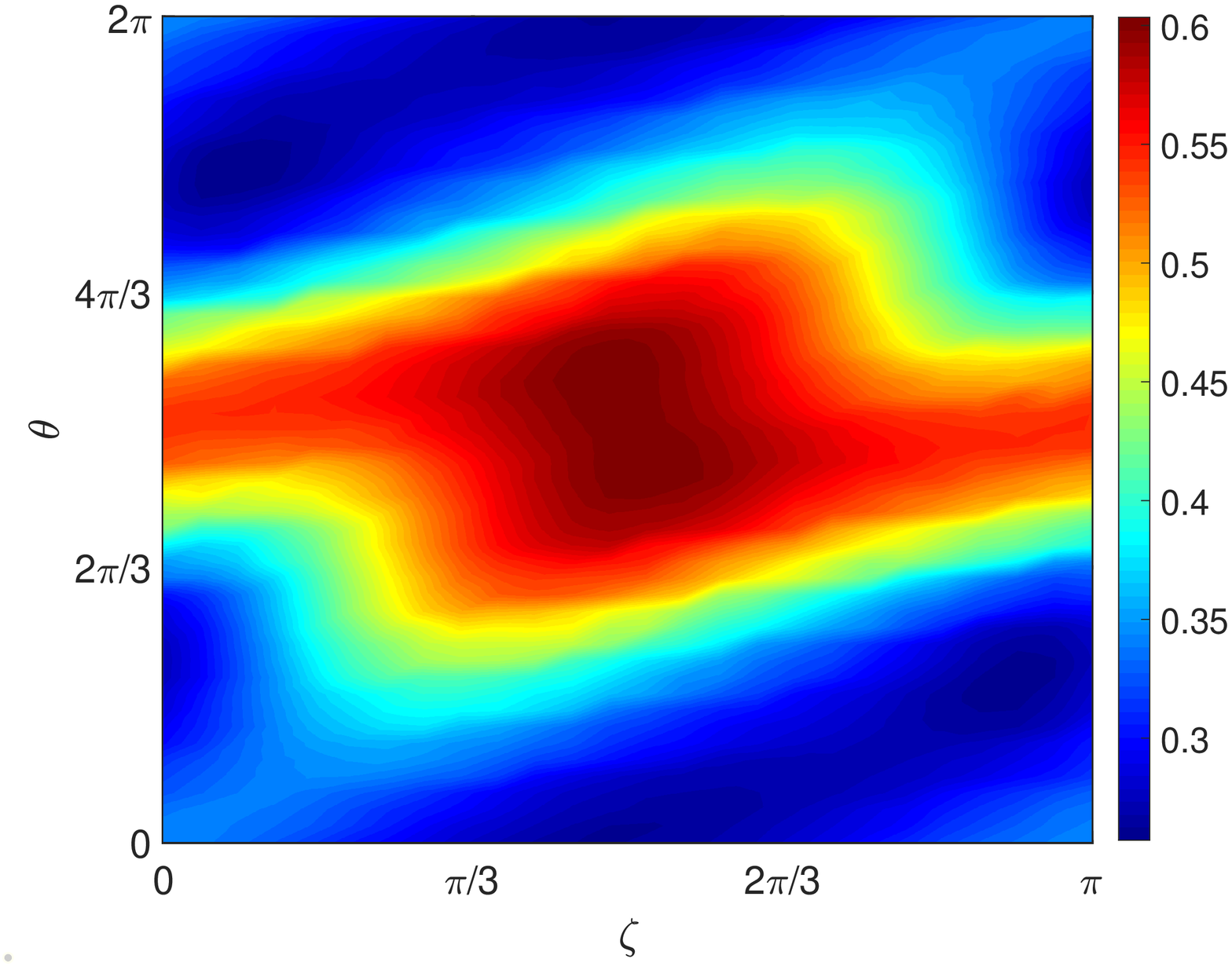}}

\caption{ $|B|$ on the boundary flux surface in the Boozer coordinate}\label{fig10}

\end{figure}

\begin{figure}
\subfigure[CNT]{
\label{fig11.a}
\includegraphics[scale=0.25]{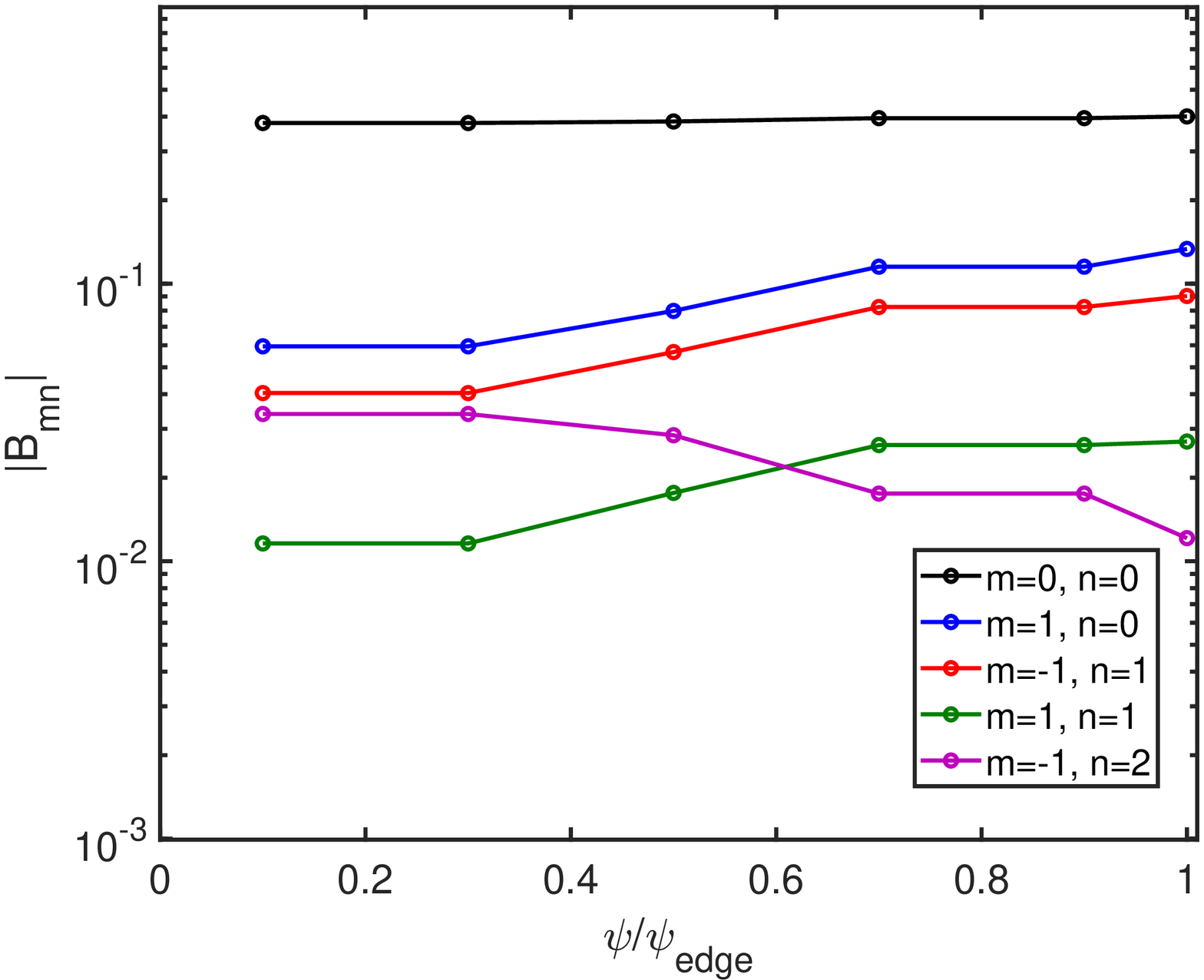}}
\subfigure[ZCS]{
\label{fig11.b}
\includegraphics[scale=0.25]{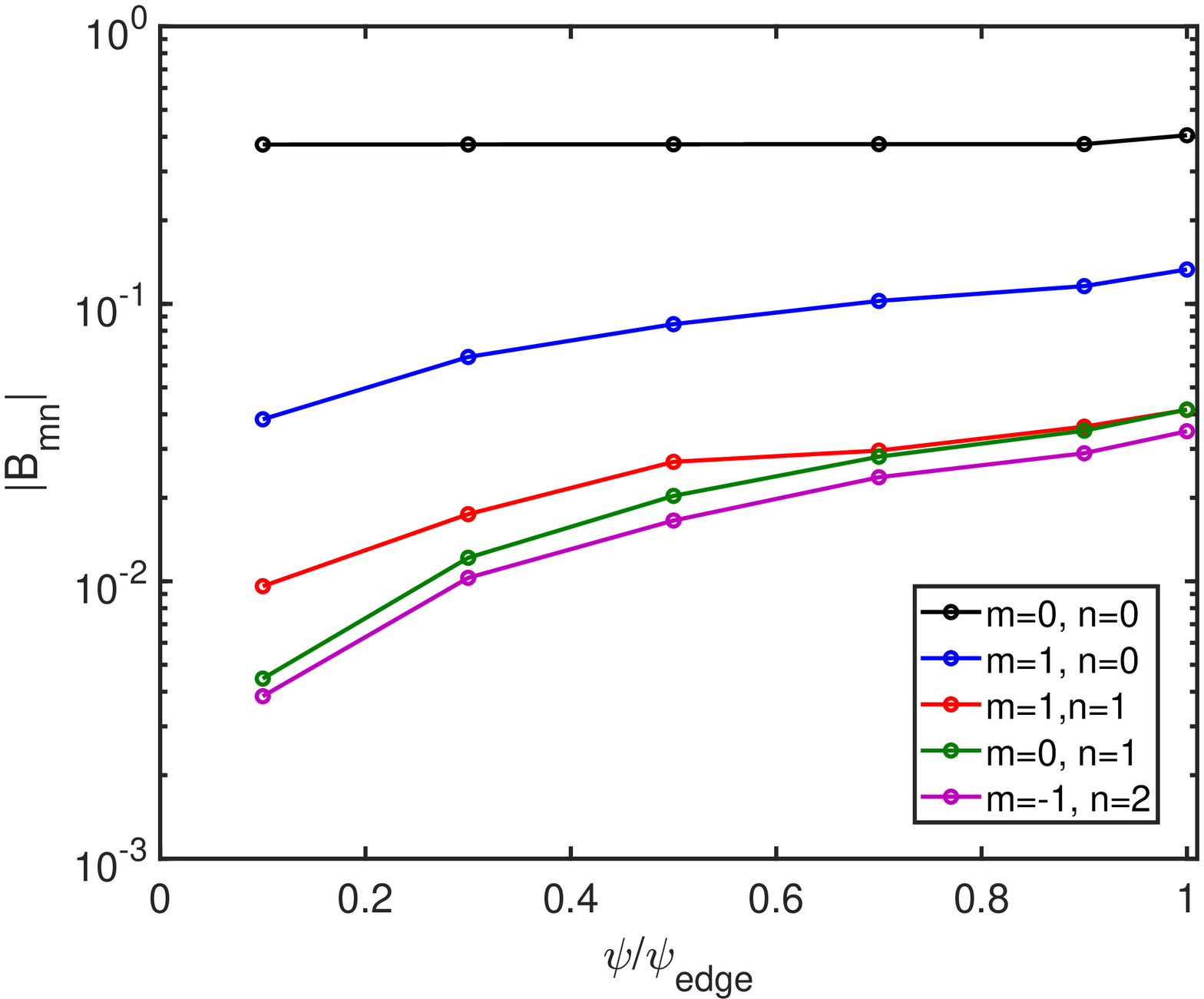}}

\caption{ Dominant $|B_{mn}|$ modes}\label{fig11}

\end{figure}

\begin{figure}

\includegraphics[scale=0.25]{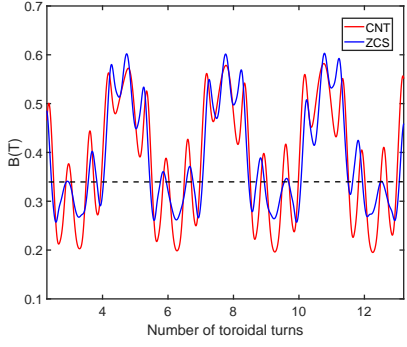}
\caption{Magnetic field strength distribution on field line}
\label{fig12}

\end{figure}

 FIG.\ref{fig10} plots the distribution of magnetic field strength on the last closed flux surface for both CNT (a) and the optimized configuration ZCS (b). We observe that the new configuration is closer to quasi-axisymmetry than CNT.  This is confirmed by Fourier spectrum of magnetic field strength shown in FIG.\ref{fig11.a} for CNT and Fig.\ref{fig11.b} for the optimized configuration. Furthermore we evaluate quasi-omnigeneity by looking at the minimums of magnetic field strength along a field line. It is known that the degree of quasi-omnigeneity can be largely measured by how close the minimums of magnetic field strength being a constant\cite{Mynick1982Class}.
In Fig.\ref{fig12}, we choose $12$ local minimums at $B \textless 0.35T$ and calculate the standard deviations of these local minimums. The result shows that the standard deviation $\delta_{ZCS}=0.05$ of ZCS is about $50\%$ of CNT's ($\delta_{CNT}=0.10$). This indicates that ZCS is much closer to quasi-omnigeneity than CNT.

\section{The finite beta effects on the effective ripple}

So far we have only considered stellarators with vacuum magnetic field and effects of finite plasma pressure have been neglected. Here we consider the effects of finite plasma beta on the helical ripple. The equilibria of ZCS at finite pressures are calculated using the VMEC code\cite{Hirshman1983Steepest}. The free boundary condition is used. The bootstrap current is calculated using SFINCS and is included in the finite beta equilibria. A few iterations between the free boundary equilibrium calculation and the bootstrap calculation are needed to obtain a converged equilibrium at finite bootstrap current. Once a converged equilibrium is obtained,
we use the NEO code to calculate the effective ripple parameter $\epsilon_{eff}^{3/2}$ at finite plasma beta.

\begin{figure}
\subfigure[Pressure profile for different plasma $\beta$]{
\label{fig13.a}
\includegraphics[scale=0.25]{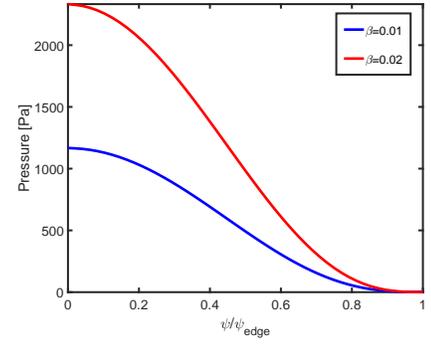}}
\subfigure[Rotation transform profile for different plasma $\beta$]{
\label{fig13.b}
\includegraphics[scale=0.25]{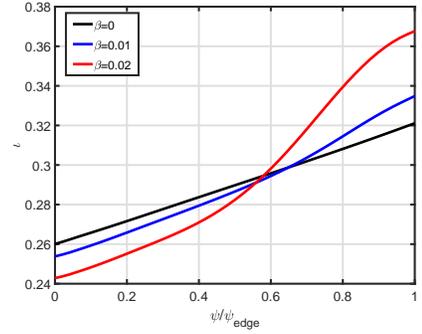}}
\subfigure[Effective ripple $\epsilon_{eff}^{3/2}$ profile for different plasma $\beta$]{
\label{fig13.c}
\includegraphics[scale=0.25]{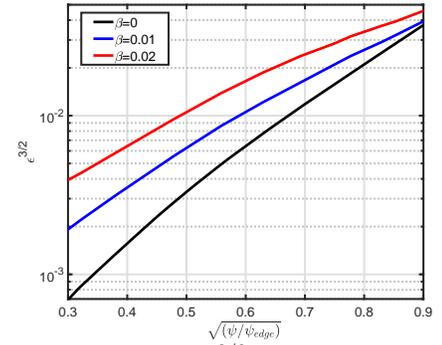}}

\caption{Pressure profile,rotation transform and effective ripple for different plasma $\beta$}\label{fig13}

\end{figure}

\begin{figure}
\subfigure{
\begin{minipage}[t]{0.25\linewidth}
\centering
\includegraphics[scale=0.25]{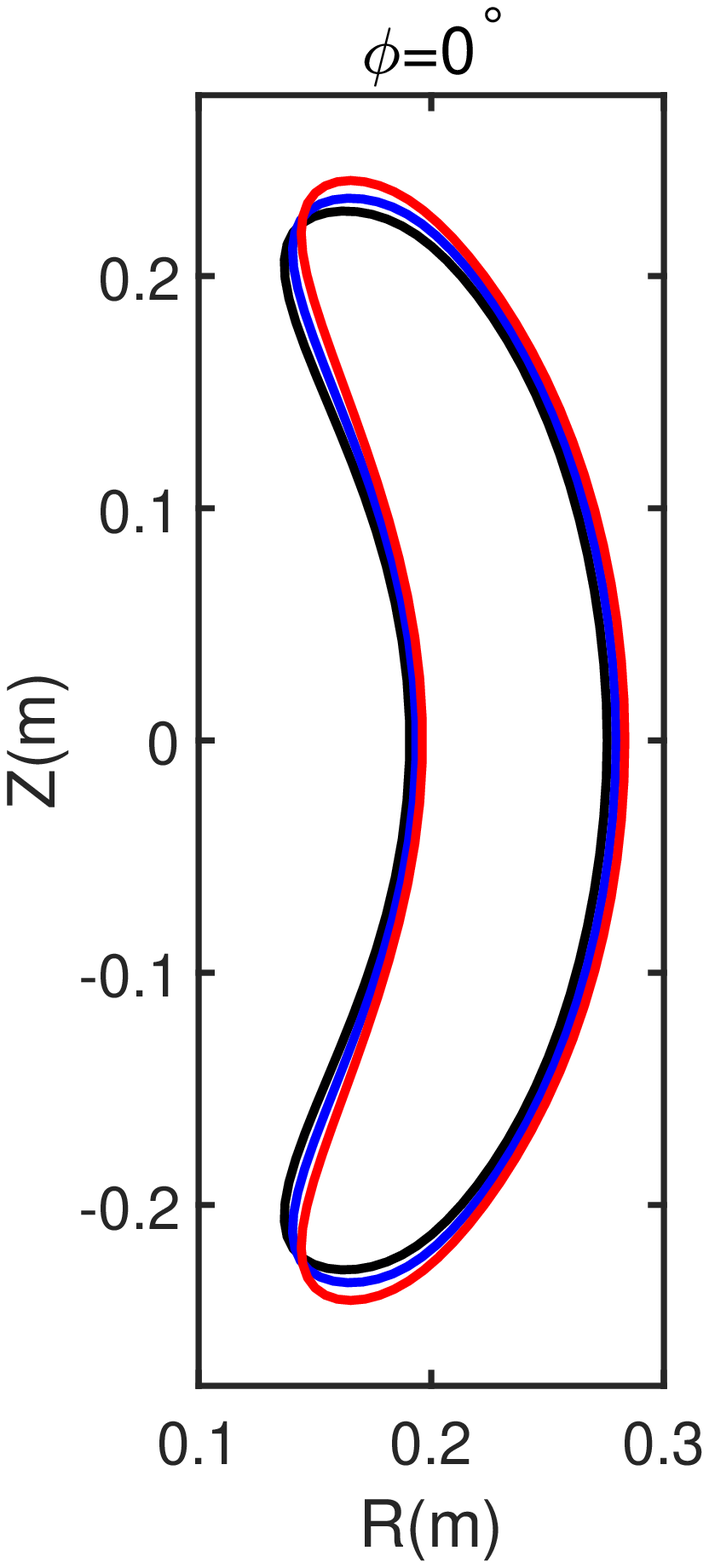}
\end{minipage}
}
\subfigure{
\begin{minipage}[t]{0.25\linewidth}
\centering
\includegraphics[scale=0.2]{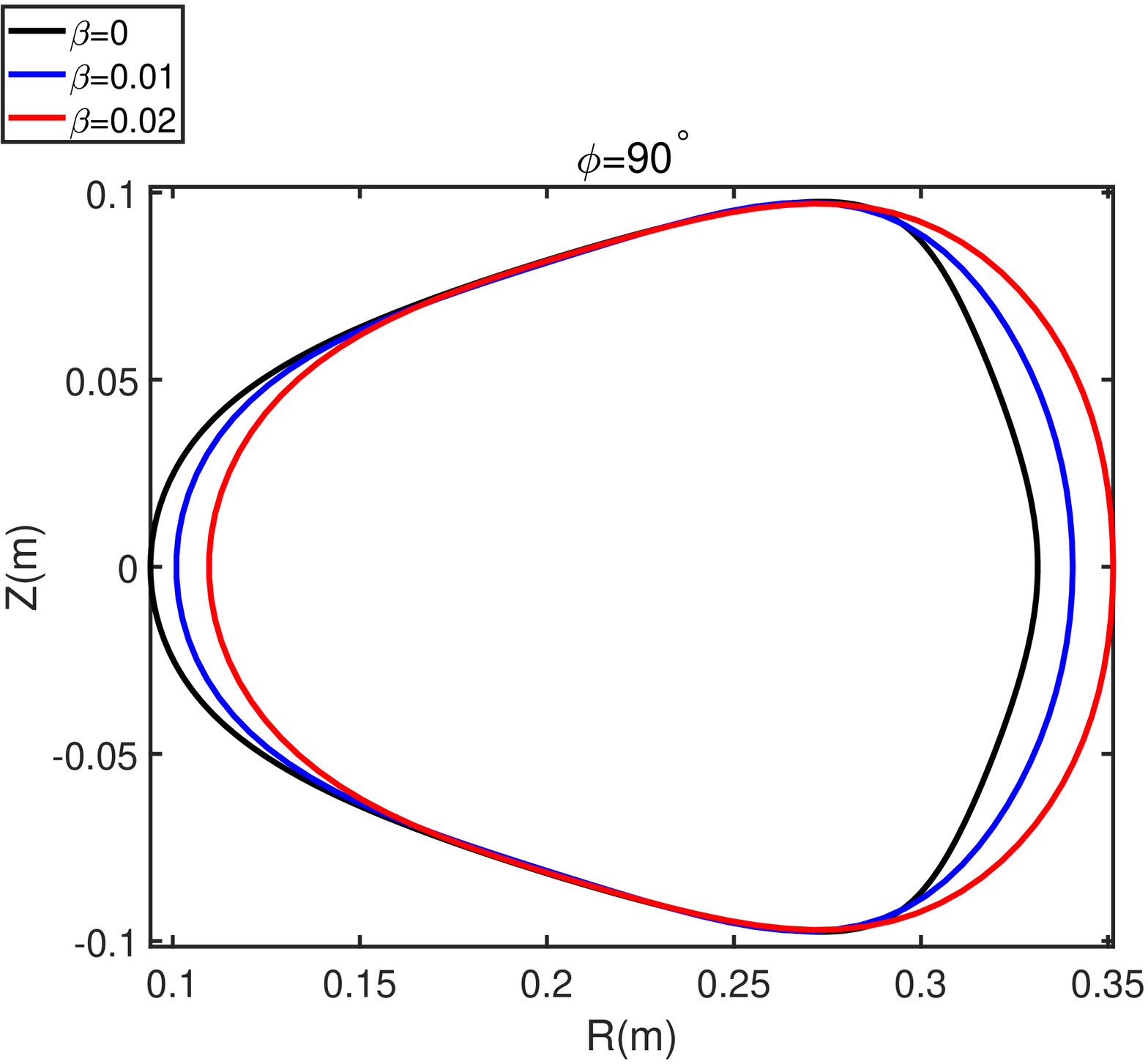}
\end{minipage}
}
\caption{Free boundary flux surfaces at $\phi=0$ and $\phi=90^{\circ}$ with same toroidal magnetic flux for different plasma $\beta$ }\label{fig14}
\end{figure}

The pressure profile is chosen to be $p=p_0(1-r^2)^3$ as shown in Fig.\ref{fig13.a} for two values of the volume-averaged plasma beta $\beta$, where $p_0$ is a constant used to control the equilibrium beta and $r$ is the square root of the normalized poloidal flux. From Fig.\ref{fig13.b} we observe that, as $\beta$ increases from zero to $2\%$, the central iota decreases slightly while the edge iota increases about $15\%$ due to bootstrap current.
Fig.\ref{fig13.c} shows that, as $\beta$ increases to $2\%$, the effective ripple parameter $\epsilon_{eff}^{3/2}$ changes little near the edge but increases by about a factor of 4 in the core.
Fig.\ref{fig14} shows the Poincare plots of the last closed magnetic surfaces for different values of plasma $\beta$. Comparing these results with those of Fig. 4, we see that the last closed surface of ZCS at $\phi=\pi/2$ with finite beta is shifted back towards the original surface of CNT. This explains, at least partly, why the effective helical ripple is enhanced at finite plasma beta. In future work the effects of finite beta will be considered in the optimization of neoclassical confinement.

\section{Conclusions}

In conclusion, a new compact stellarator with simple coils and good neoclassical confinement has been designed. The magnetic field of the new stellarator is generated by only four planar coils including two interlocking coils of elliptical shape and two circular poloidal field coils. The neoclassical optimized configuration was obtained by a global minimization of the effective helical ripple directly from the shape of the two interlocking coils. The optimized compact stellarator has very low level of effective ripple in the plasma core implying excellent neoclassical confinement. The results of the drift-kinetic code SFINCS show that the particle flux of the new configuration is one order of magnitude lower than CNT's in the core. Future work will consider optimization of neoclassical confinement and MHD stability at finite beta.

\section*{Acknowledgement}

We are indebted to Dr. Caoxiang Zhu for useful discussions and for his help in use of the stellarator optimization code STELLOPT. This work was funded by the start-up funding of Zhejiang University for one of the authors (Prof. Guoyong Fu).

\bibliographystyle{unsrt}
\bibliography{2021_6_10}

\end{document}